\documentclass[final,3p,times,twocolumn]{elsarticle}
\usepackage{amsmath,amssymb,amsfonts}
\usepackage{graphicx}
\usepackage{threeparttable}
\usepackage{textcomp}
\usepackage[linesnumbered,ruled,vlined]{algorithm2e}%[ruled,vlined]{  
\usepackage{subfigure}
\usepackage{stfloats}
\usepackage{algpseudocode}  
\usepackage{amsmath}
\usepackage{booktabs}
\usepackage{subcaption}
\usepackage{amssymb}
\journal{Soft Computing}

\begin{document}

\begin{frontmatter}


% \author{Name\corref{cor1}\fnref{label2}}
% \affiliation{organization={},

\title{Influence Maximization Considering Influence, Cost and Time}


\author[label1,label2]{Mingyang Feng}
\ead{11856010@mail.sustech.edu.cn}
\author[label1]{Qi Zhao}
\ead{zhaoq@sustech.edu.cn}
\author[label2]{Shan He}
\ead{s.he@cs.bham.ac.uk}
\author[label1]{Yuhui Shi \corref{cor1}}
\ead{shiyh@sustech.edu.cn}
\cortext[cor1]{Yuhui Shi is the corresponding author}

\affiliation[label1]{organization={Department of Computer Science and Engineering, Southern University of Science and Technology},%Department and Organization
            %addressline={}, 
            city={Shenzhen},
            postcode={518055}, 
            %state={},
            country={China}}
\affiliation[label2]{organization={School of Computer Science, The University of Birmingham},%Department and Organization
            %addressline={}, 
            city={Birmingham},
            postcode={B15 2TT}, 
            %state={},
            country={UK}}
\begin{abstract}

Influence maximization has been studied for social network analysis, such as viral marketing (advertising), rumor prevention, and opinion leader identification. However, most studies neglect the interplay between influence spread, cost efficiency, and temporal urgency. In practical scenarios such as viral marketing and information campaigns, jointly optimizing Influence, Cost, and Time is essential, yet remaining largely unaddressed in current literature. To bridge the gap, this paper proposes a new multi-objective influence maximization problem that simultaneously optimizes influence, cost, and time. We show the intuitive and empirical evidence to prove the feasibility and necessity of this multi-objective problem. We also develop an evolutionary variable-length search algorithm that can effectively search for optimal node combinations. The proposed EVEA algorithm outperforms all baselines, achieving up to 19.3\% higher hypervolume and 25–40\% faster convergence across four real-world networks, while maintaining a diverse and balanced Pareto front among influence, cost, and time objectives. We release the source code on GitHub: https://github.com/fmyzckj.

\end{abstract}




\end{frontmatter}


\section{Introduction}
\label{}
Influence Maximization (IM) is a foundational problem in social network analysis, focusing on selecting a small set of seed users to maximize information spread across the network via mechanisms like social influence or viral propagation. Pioneered by Kempe et al. \cite{kempe2003maximizing} with the Independent Cascade (IC) and Linear Threshold (LT) models, IM has become pivotal in both academia and practice. Its applications span viral marketing, where businesses leverage influential users to amplify product adoption \cite{chen2010scalable}; public health, aiding in identifying super-spreaders for targeted interventions \cite{pastor2015epidemic}; and social recommendation systems, enhancing engagement through peer influence \cite{li2017towards}.

\begin{figure*}[ht]
    \centering
    \subfigure[]{
    \label{unreasonable}
    \includegraphics[height=5.5cm,width=7.5cm]{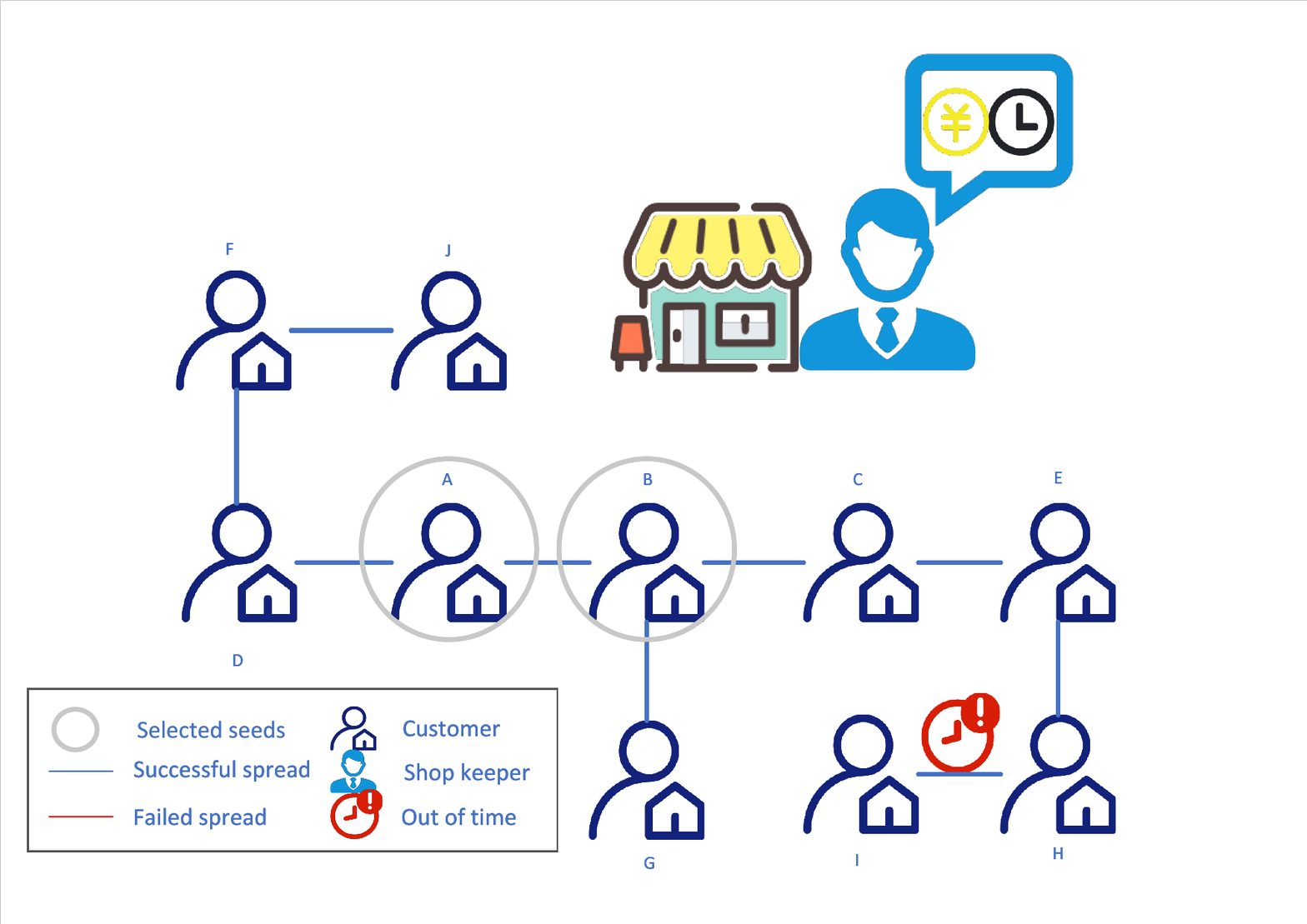}
    }
    \subfigure[]{
    \label{reasonable}
    \includegraphics[height=5.5cm,width=7.5cm]{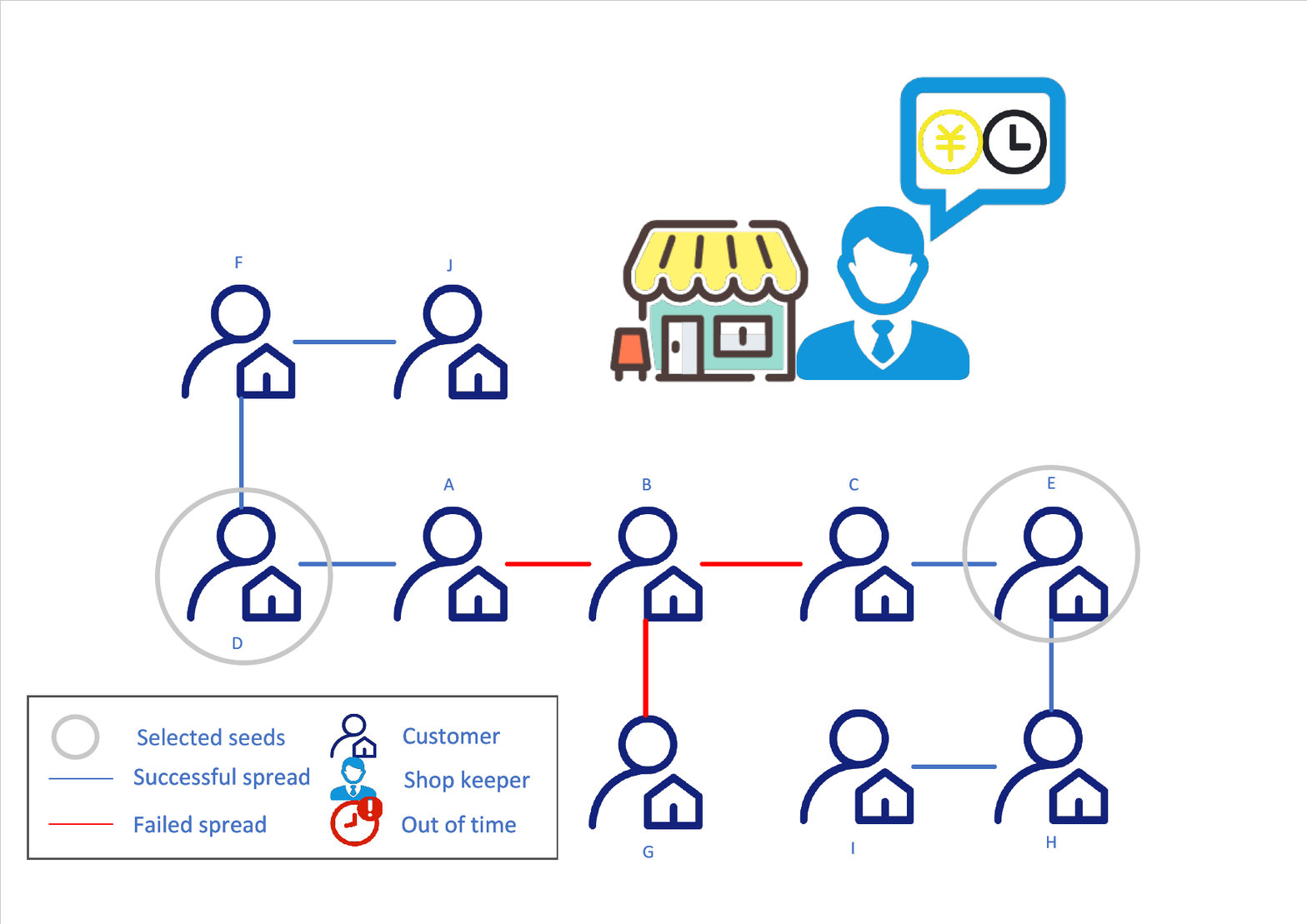}
    }
    \caption{supermarket}
    \label{fig:my_label1}
\end{figure*}

In this paper, we formalize a novel multi-objective influence maximization problem that simultaneously optimizes three conflicting criteria: influence coverage, activation cost, and propagation time (termed as IM-ICT). To demonstrate the necessity of this formulation, consider the scenario in Figure~\ref{fig:my_label1} where a supermarket promotes a time-sensitive sale through a social network of 10 users (A-J). The shopkeeper must select two seed users under constrained budgets and promotional deadlines. We define node activation costs proportional to their degrees (reflecting influencer pricing) and propagation time as the longest path from seeds to influenced nodes. Figure~\ref{unreasonable} illustrates a maximal influence strategy selecting high-degree seeds A (degree=2) and B (degree=3), achieving full network coverage at a total cost of 5 units. However, the critical path B→C→H→I requires 4 time steps, exceeding the permissible 3-step threshold. By contrast, Figure~\ref{reasonable} demonstrates a Pareto-optimal solution choosing seeds D (degree=2) and E (degree=2), which reduces both cost (4 units) and time (2 steps) while sacrificing partial coverage (nodes B and G remain unaffected). This contrast highlights fundamental trade-offs between the objectives: high influence demands premium seeds and extended propagation horizons, whereas cost-time efficiency necessitates strategic coverage compromises. Beyond commercial promotions, our IM-ICT formulation also enables nuanced decision-making in public health (e.g., balancing vaccine distribution costs, immunization speed, and population coverage) and cybersecurity (e.g., minimizing containment costs while maximizing malware eradication speed and scope).

The optimization of the proposed IM-ICT problem incurs two algorithmic challenges. First, we rigorously prove the proposed IM-ICT's NP-hardness via a polynomial-time reduction from the Set Cover problem. In addition to the NP-hard challenge, the length of solutions requires variability to enable the algorithm to naturally evolve seed sets with different sizes to suit the diverse optimization pressures from conflicting objectives, leading to a more complete and expressive Pareto front. To address these challenge, we propose an Embedding-aligned Variable-length Evolutionary Algorithm (EVEA), which introduces a graph embedding-driven crossover operator and enables variable-length search by aligning the nodes with embedding vectors. 

In summary, our contributions are:
\begin{enumerate}
    \item The new IM-ICT problem formulation with real-world constraints. Unlike prior single- or bi-objective formulations, our model explicitly integrates propagation time as the longest path length from seeds to influenced nodes, reflecting real-world urgency requirements (e.g., limited-duration sales or epidemic containment). We provide both theoretical justification (NP-hardness proof) and empirical evidence (case studies in Figures 1–2) to validate the necessity of this new formulation.
    \item The Embedding-aligned Variable-length Evolutionary Algorithm for solving IM-ICT. The proposed algorithm's core innovation lies in its crossover operator, which aligns parent solutions using node embedding vectors. This enables the recombination of seed sets with varying lengths while preserving topological semantics. Such operations enhance search space exploration and accelerate convergence.
    \item Comprehensive Empirical Validation. Experiments on real-world networks (Twitter, DBLP) demonstrate EVEA’s superiority over state-of-the-art multi-objective optimizers.

\end{enumerate}

The remainder of this paper is organized as follows: Section 2 reviews related work on IM and multi-objective optimization. Section 3 formally defines the proposed IM-ICT problem and analyze its complexity. Section 4 details EVEA’s architecture, focusing on the embedding-aligned crossover mechanism. Section 5 presents experimental results, and Section 6 concludes with future directions.

\section{Related Works}

This section provides a comprehensive overview of the IM problem, its foundational formulations, and recent advances in both single-objective and multi-objective settings. We first introduce the problem definition and diffusion models, followed by a discussion of classical single-objective approaches, and finally review the state-of-the-art in multi-objective influence maximization.

\subsection{Problem Formulation: Influence Maximization in Networks}

IM is a fundamental problem in network science, originally proposed by Kempe et al. \cite{kempe2003maximizing}, which aims to select a small set of seed nodes within a social network to maximize the expected number of activated nodes under a chosen diffusion model.

Formally, consider a directed graph \( G = (V, E) \), where \( V \) represents the set of nodes and \( E \) the set of edges. Each edge \( (u,v) \) has an associated propagation probability \( p_{uv} \in [0,1] \). The goal is to select a seed set \( S \subseteq V \) of size \( k \) that maximizes the expected influence spread, defined as:
\begin{equation}
S^* = \arg\max_{S \subseteq V, |S|=k} \sigma(S)
\end{equation}

where \( \sigma(S) \) denotes the expected number of activated nodes when the diffusion process starts from \( S \).

The most widely used diffusion models in IM are the \textit{IC model} and the \textit{LT model}. In the IC model, diffusion proceeds in discrete time steps. Each active node \( u \) has a single chance to activate each of its inactive neighbors \( v \), with probability \( p_{uv} \). The process continues until no further activations occur. The expected influence spread under the IC model is formally defined as:

\begin{equation}
\sigma(S) = \mathbb{E} \left[ \sum_{v \in V} \mathbb{I}(v \text{ is activated}) \right]
\end{equation}
where \( \mathbb{I}(\cdot) \) is the indicator function.

In the LT model, each node \( v \) is assigned a threshold \( \theta_v \in [0,1] \). A node becomes active when the sum of incoming weights from its active neighbors exceeds its threshold. The LT model captures cumulative influence effects, whereas the IC model emphasizes independent, probabilistic activations.

To address time-sensitive scenarios, recent extensions like the \textit{Latency-Aware Independent Cascade (LAIC) model} incorporate temporal delays into the diffusion process. In this model, each activation attempt from node \( u \) to node \( v \) introduces a delay \( \delta t \) sampled from a distribution \( P_u(\delta t) \). The propagation time \( T(S) \) of a seed set \( S \) can be expressed as the maximum activation time across all influenced nodes:
\begin{equation}
T(S) = \max_{v \in V} \left\{ \min_{u \in S} \left( t(u) + d(u,v) \cdot \delta t \right) \right\}
\end{equation}
where \( d(u,v) \) is the shortest path length between \( u \) and \( v \) in the network, and \( t(u) \) is the activation time of the seed node \( u \) (typically \( t(u) = 0 \)).

The Influence Maximization problem is known to be NP-hard under both the IC and LT models, and approximating the expected spread \( \sigma(S) \) is \#P-hard. Consequently, solving IM requires either approximation algorithms or heuristic approaches that balance accuracy and efficiency.

\subsection{Influence Maximization: Single-Objective Approaches}

Since its introduction, Influence Maximization has been predominantly studied as a single-objective optimization problem, focusing on maximizing the spread of influence. Kempe et al. \cite{kempe2003maximizing} proposed a greedy algorithm that achieves a \((1-1/e)\)-approximation guarantee under submodular diffusion models, but the method’s high computational cost---due to the need for repeated Monte Carlo simulations---prompted subsequent research into more efficient techniques. Leskovec et al. \cite{leskovec2007cost} introduced the CELF algorithm, which improves greedy selection efficiency by exploiting submodularity properties to minimize redundant computations. Further advances include sketch-based methods such as TIM and IMM \cite{tang2014influence,tang2015influence}, which approximate influence spread via reverse reachable (RR) sets, significantly reducing computation time while maintaining high solution quality. 

Beyond exact or approximate methods, heuristic strategies like degree centrality, PageRank, and community detection have been widely explored for their scalability \cite{PRDiscount2015, PRDD2021, Aghaee2021IMPTC, Umrawal2022CommunityAwareIM, Jalayer2021GTaCB}, albeit at the expense of accuracy and theoretical guarantees. In parallel, evolutionary and swarm-based algorithms---such as particle swarm optimization (PSO) \cite{gong2016influence, yang2017influence} and biogeography-based optimization (BBO) \cite{Salehi2020KCBBO, De2020MOBBO} ---have been proposed to balance influence spread and computational cost by exploring large solution spaces using population-based search strategies. These methods have demonstrated success in handling larger networks but remain limited by their focus on a single objective: maximizing the number of influenced nodes, without considering factors like seed acquisition costs, propagation delays, or fairness constraints that often arise in real-world applications.

\subsection{Multi-Objective Influence Maximization: Recent Advances}

Recognizing the limitations of single-objective formulations, researchers have increasingly focused on Multi-Objective Influence Maximization, where influence spread is optimized alongside other conflicting objectives such as seed cost, propagation latency, and fairness \cite{chen2012time, Tsang2019GroupFairIM, Ali2021FairTCIM, Gershtein2021MOIM, Cunegatti2024MOEIM}. The trade-off between influence and cost has been a primary focus, as high-impact seed nodes---often those with high degree or centrality---are typically associated with higher acquisition costs in practical scenarios. Several studies have proposed algorithms to address this trade-off. Wang and Zhang \cite{Wang2023MOCSA_IM} introduced a Multi-Objective Crow Search Algorithm (MOCSA) that simultaneously maximizes influence spread and minimizes seed costs, demonstrating superior performance across real-world networks. Yang and Liu \cite{yang2017influence} proposed a multi-objective particle swarm optimization framework for influence maximization under budget constraints, offering diverse Pareto-optimal solutions that balance influence and cost. De et al. \cite{De2020MOBBO} developed a biogeography-based multi-objective algorithm that optimizes influence spread while reducing budget expenditure, and Gershtein et al. \cite{Gershtein2021MOIM} presented IM-Balanced, a system allowing users to specify trade-offs between objectives, enabling tailored solutions for different applications.

Another major research direction in MOIM is the inclusion of propagation latency as an optimization objective. In time-sensitive scenarios, such as viral marketing campaigns or emergency information dissemination, the speed at which influence spreads is as critical as the extent of coverage. Chen et al. \cite{chen2012time} were among the first to explore time-critical influence maximization, incorporating propagation delays into the IC and LT models. Li et al.~(2020) extended this line of work by proposing a latency-aware influence maximization model that optimizes both influence and diffusion speed. Wang et al. \cite{Wang2024IMOPSO} developed a multi-objective evolutionary algorithm that minimizes both latency and seed cost while maintaining competitive influence spread. 

Fairness in influence propagation has also gained attention in recent years. In heterogeneous networks, ensuring equitable influence distribution across different groups---such as demographic segments, geographic regions, or interest communities---is crucial for ethical and practical reasons. Tsang et al. \cite{Tsang2019GroupFairIM} introduced a group fairness-aware influence maximization framework that ensures balanced coverage across diverse user groups. Staib and Jegelka \cite{Staib2018DRO} studied constrained submodular maximization problems, applying fairness constraints to influence maximization settings. Farnad et al. \cite{Farnad2020FairIM} designed algorithms that jointly optimize for influence spread and fairness metrics, such as equal opportunity, while Rahmattalabi et al. \cite{Rahmattalabi2021FairIM} proposed fairness-constrained formulations to address exposure disparities across communities.

Beyond these primary objectives, recent studies have explored multi-objective influence maximization in dynamic networks, competitive settings, and under additional constraints such as privacy, energy efficiency, and physical distance \cite{Zhu2021IMPR, Cunegatti2024MOEIM, Genetti2025hnMOEA, Huang2022ClosedIM, Tong2015AdaptiveIM}. Despite these advances, challenges remain in designing scalable algorithms that effectively balance multiple objectives while adapting to real-world complexities.

\section{Proposed IM-ICT Problem}
\subsection{Objectives}
Building upon the conventional IM definition in Section 2.1, we formalize our proposed model with three optimization objectives:

\textbf{Objective 1: Influence Spread Maximization:}
Consistent with traditional IM, our primary objective aims to maximize the number of influenced nodes in the network. This can be expressed as:
\begin{equation}
f_{1}=\sigma(S)=\mathbb{E} \left[\sum_{v \in V}\mathbb{I}_{(v \ activated)} \right]
\label{influence}
\end{equation}
where node $v$ becomes activated through multi-round diffusion if any predecessor $u \in \mathcal{N}^{-}(v)$ successfully influences it through edge $(u,v)$.

\textbf{Objective 2: Seed Cost Minimization:}
Under the IM framework that assumes activation of $k$ seed nodes at initial time, we introduce a cost set $C$ representing activation expenses for each node (extending the network definition to $G=(V,E,C)$). The secondary objective minimizes the total seed acquisition cost:
\begin{equation}
   f_{2}= B(S) = \sum_{v_{i} \in S}c_i
\label{cost}
\end{equation}
where $c_{i}$ represents the activation cost of node $v_i$ in seed set $S$, and $B(S)$ denotes the total budget required for seed set $S$.

\textbf{Objective 3: Propagation Latency (Time) Minimization:}
Under the Latency-Aware Independent Cascade (LAIC) model, we define propagation latency as the latest activation time among all influenced nodes (To simplify this, we call this objective time in this paper):
\begin{equation}
    f_{3}=T(S)=\max_{v_{i} \in S, v_{j} \in V}\left[t_i+\sum_{k=1}^{d(v_{i},v_{j})}\delta_{t_{k}}   \right ]  
\label{time}
\end{equation}
where $t_i$ indicates the initial activation time of seed $v_{i}$ (typically $t_{i}=0$), $d(v_{i},v_{j}) $represents the shortest path length from $v_{i}$ to $v_{j}$, and $\delta_{t_{k}}$ denotes the activation delay at the $k$-th hop sampled from distribution $P^{lat}_u(\delta_t)$.

As analyzed in Section I, inherent conflicts exist among these objectives:
\begin{enumerate}
    \item Influence-Cost Conflict: Maximizing influence spread typically requires selecting high-impact opinion leaders as seeds, which substantially increases activation costs.
    \item Influence-Time Tradeoff: Broad coverage often necessitates longer propagation paths, thereby increasing time delays.
    \item Cost-Time Dilemma: Cost minimization strategies tend to select peripheral nodes, which may compromise temporal effectiveness due to their marginal network positions.
\end{enumerate}

\subsection{IM-ICT}
Therefore, the IM-ICT problem can be formulated as follows:

\begin{equation}
\begin{aligned}
    \max F &= (f_1, -f_2, -f_3) \\
    \text{s.t.} \quad & S \subseteq V \ \&\ |S| \geq 1 \\
    & c_i \geq 0, \quad \forall v_i \in V \\
    & p_{uv} \in [0, 1] \\
    & \delta_{t} \sim P^{lat}_u(\delta_t) \ \&\ \delta_{t} \geq 0
\end{aligned}
\label{three obj}
\end{equation}
where $p_{uv}$ denotes the propagation probability from node $u$ to $v$.

% In this section, we will first introduce preliminary knowledge of IM problem. Including the definition of the problem and the introduction of the propagation model. Then we introduce the proposed three-objectives problem model. 

% In a network $G=(V, E)$, where $V$ represents the nodes set in the network and $E$ represents the edges set. The aim of the IM problem is to find a set of $S=\{v_1,,v_2,...,v_k\}$ which contains $k$ nodes ($k$ is an integer and $k<|V|$) and the influence of S is maximal under a certain influence spread model.

% Summarizing the above three objectives, we give the definitions and formulas of the multi-objectives as follows:

% \textbf{Definition}: Given a network $G=(V,E,C)$, a range of $[k_{min},k_{max}]$ and the latency distribution $P^{lat}_u(\delta_t)$. The aim of this problem is to find a set of nodes which contains at most $k_{max}$ nodes and at least $k_{min}$ nodes. These nodes satisfy the follow three objectives:

\section{Proposed Embedding-aligned Variable-length Evolutionary Algorithm}
In this section, we present the design and working principles of our proposed algorithm, the Embedding-aligned Variable-length Evolutionary Algorithm (EVEA), developed to tackle the IM-ICT problem under multiple conflicting criteria.

\subsection{Overall framework}
EVEA is a population-based evolutionary algorithm that optimizes influence maximization solutions by balancing competing objectives including influence spread, activation cost, and time. Unlike traditional approaches, EVEA adopts a variable-length representation for candidate solutions, allowing the number of seed nodes in each solution to adaptively evolve during the search process. This flexibility enables the algorithm to explore a diverse range of trade-offs and achieve a well-distributed Pareto front. Additionally, the algorithm integrates embedding-based knowledge to enhance solution recombination, further promoting efficiency and convergence in the multi-objective setting.

The pseudo-code of EVEA is shown in Algorithm~\ref{alg:naga-ii}. The algorithm starts with the initialization of a population of candidate solutions, each representing a distinct seed set. At each generation, a tournament selection is performed to select parent solutions. The selected parents undergo an embedding-aligned crossover to produce offspring, followed by a variable-length mutation operation that modifies the seed set size and composition. The combined parent and offspring populations are then subjected to a non-dominated sorting selection (as in NSGA-II \cite{deb2002fast}) to maintain Pareto-optimal solutions. The algorithm iterates until a termination condition is met (e.g., a fixed number of generations). In the following, we detail the core innovations of the algorithm, i.e., the embedding aligned crossover and variable-length mutation, respectively.

\begin{algorithm}[t]
\caption{Embedding-aligned Variable-length Evolutionary Algorithm}
\label{alg:naga-ii}
\KwIn{$G = (V,E,C)$, Generation $G_{max}$, Network size $N$}
\KwOut{Pareto front $p_f$ of population $P_t$}
$Generation \ counter \ G_t = 0$

$P^G \gets Initialization()$

\While{stopping criterion not met}{
$P^G \gets Tournament\_Selection(P^G)$

$Q \gets Embedding-Aligned\_Crossover(Q)$ 

$Q \gets Variable-Length\_Mutation(Q)$

$P_G \gets Environmental\_Selection(P_G \cup Q)$

$G_t = G_t+1$

}

$P_t = P_G$

$p_f \gets P_t$
\end{algorithm}

\subsection{Embedding Aligned Crossover}
The core innovation of EVEA lies in its embedding-aligned crossover operator (Algorithm~\ref{alg:crossover}). Unlike traditional crossover operators that require equal-length representations, this operator aligns nodes from two parent solutions using their graph embedding vectors (e.g., derived from Node2Vec \cite{grover2016node2vec} or DeepWalk \cite{perozzi2014deepwalk}, in this paper, we choose to use Node2Vec). Specifically, given two parent solutions $S_1$ and $S_2$, we compute the pairwise Euclidean distances between their nodes' embeddings. Nodes with the smallest distances are paired, and each pair has a probability $p_c$ of swapping their membership across solutions.

\begin{figure}[ht]  % 使用figure环境而非figure*实现单栏排版
    \centering
    \subfigure[]{
    \includegraphics[width=0.8\linewidth,height=1.5cm]{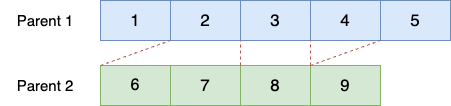}
    }\\
    \vspace{0.2cm}  % 调整图片间距
    
    \subfigure[]{
    \includegraphics[width=0.8\linewidth,height=3.5cm]{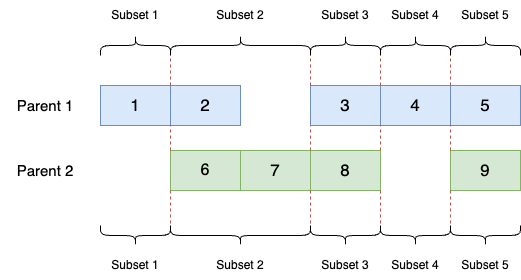}
    }\\
    \vspace{0.2cm}
    
    \subfigure[]{
    \includegraphics[width=0.8\linewidth,height=3.5cm]{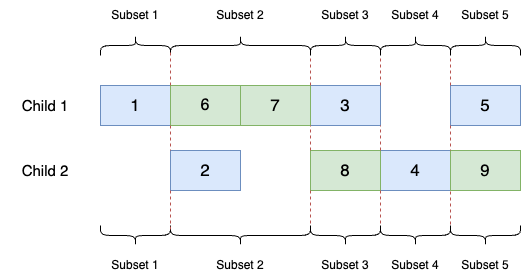}
    }\\
    \vspace{0.2cm}
    
    \subfigure[]{
    \includegraphics[width=0.8\linewidth,height=1.5cm]{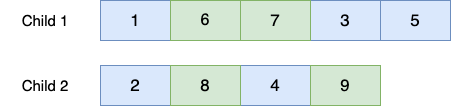}
    }
    \caption{Embedded crossover}
    \label{fig:crossover}
\end{figure}

To further clarify the embedding-aligned crossover process, we provide a visual illustration in Figure~\ref{fig:crossover}, which demonstrates the key steps of the operator. In this figure, part (a) depicts two parent solutions represented in a two-dimensional embedding space, where nodes are positioned according to their topological and semantic similarity. Part (b) shows the alignment process, where nodes from different parents are paired based on their proximity in the embedding space. Part (c) illustrates the actual crossover operation, where aligned node pairs are exchanged with a certain probability, facilitating the recombination of seed sets. Finally, part (d) presents the offspring solutions generated through this crossover, highlighting how the embedding-guided approach preserves structural semantics while enabling diversity in solution construction.

By leveraging the embedding-based alignment, this crossover strategy allows EVEA to recombine solutions of varying seed set sizes effectively, avoiding the limitations of index-based or fixed-length operators. This design enhances the algorithm’s ability to explore the solution space meaningfully, promoting better convergence toward high-quality Pareto fronts.

\begin{algorithm}[t]
\caption{Embedding Aligned Crossover}
\label{alg:crossover}
\KwIn{Parents $S_1$,$S_2$,Embedding $EM$, Crossover Rate $p_c$}
\KwOut{Children $C_1$,$C_2$}
Align nodes in  $S_1$, $S_2$ by embedding (Euclidean Distance)

For each paired nodes, exchange nodes with probability $p_c$

Output $C_1$, $C_2$

\end{algorithm}

\subsection{Variable Length Mutation}
To further promote diversity and adaptability, EVEA incorporates a variable-length mutation operator (Algorithm~\ref{alg:mutation}). At each mutation step, one of the following strategies is randomly selected with equal probability:

\begin{itemize}
    \item Add: A random node not currently in the solution is added to the seed set.
    \item Delete: A random node from the solution is removed, reducing the seed set size.
    \item Replace: A random node in the solution is swapped with another node outside the solution.
\end{itemize}

These operations enable EVEA to flexibly adjust the size of the seed set, allowing the algorithm to navigate complex trade-offs and better adapt to the multi-objective nature of the problem.

\begin{algorithm}[t]
\caption{Variable length Mutation}
\label{alg:mutation}
\KwIn{Solution $S$, Mutation rate$p_m$}
\KwOut{Solution $S^*$}

$Strategy \gets Random(\{add, delete, replace\})$

\If{$Strategy = add$}{
$v \gets G-S$

$S^* \gets S \cup \{v\} $
}
\ElseIf{$Strategy = delete$}{
$v \gets S$

$S^* \gets S \backslash \{v\} $
}
\ElseIf{$Strategy = replace$}{
$v_G \gets G$

$v_S \gets S$

$S^* \gets S \cup \{v_G\}\backslash \{v\} $
}

Output $S^*$

\end{algorithm}

\subsection{Time Complexity Analysis}
The time complexity of the proposed algorithm is primarily determined by the multi-objective selection mechanism, which ensures the preservation of high-quality and diverse solutions across generations. In each iteration, the algorithm evaluates a population of $N$ candidate solutions and applies a selection strategy based on Pareto dominance relations and crowding distance calculations to maintain a well-distributed Pareto front. This selection step has a time complexity of $O(N^2)$.

Additionally, the embedding-aligned crossover operator involves at most $\min (k_1,k_2)$ pairwise distance computations between seed nodes in the parent solutions, where $k_1$ and $k_2$ are the lengths of the respective parent solutions. Since $k_1,k_2\ll|V|$ (with $|V|$ being the number of nodes in the network), this step introduces negligible overhead relative to the overall selection process.

Therefore, the overall time complexity of each generation is $O(N^2)$, ensuring scalability and efficiency while maintaining the algorithm's ability to effectively balance trade-offs between influence spread, activation cost, and time.

% \section{Experimental Setting and Results}
% In this section, we give the experiment settings including datasets, parameter setting and algorithms for comparison. Then we show the results including conflicts verifying and algorithm comparison.
% \subsection{Experimental Settings}

% \subsection{Experimental Results}

% \subsubsection{Verify the conflicted relationship between the three objectives}

% \subsubsection{Compare with other algorithms}

% \subsection{Convergence of EVEA}
\begin{figure*}[]
     \centering  
     \subfigure[Facebook]{
         \centering
         \includegraphics[width=0.45\textwidth]{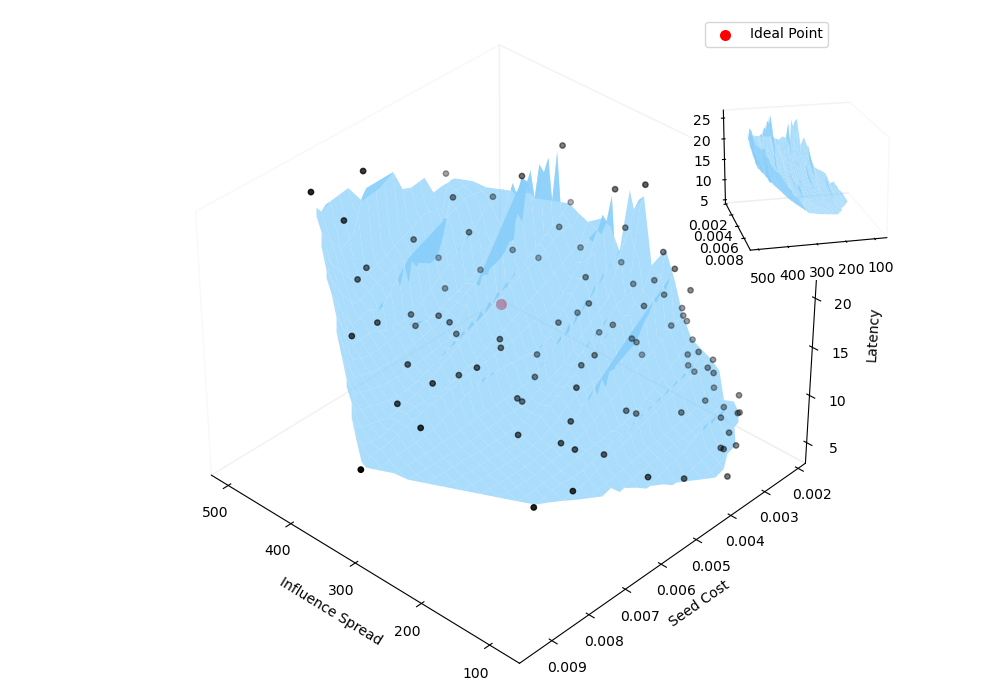}
         \label{paretofig:4000}}
     ~
     \subfigure[Grqc]{
         \centering
         \includegraphics[width=0.45\textwidth]{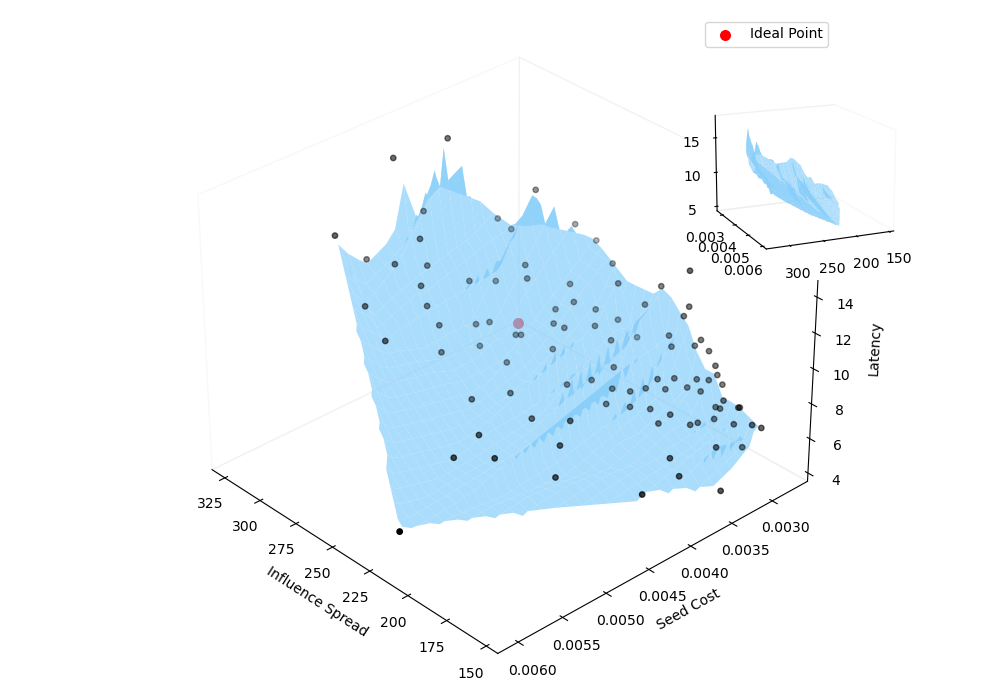}
         %\caption{5k}
         \label{paretofig:5000}}
     ~
     
     \subfigure[Gnutella]{
         \centering
         \includegraphics[width=0.45\textwidth]{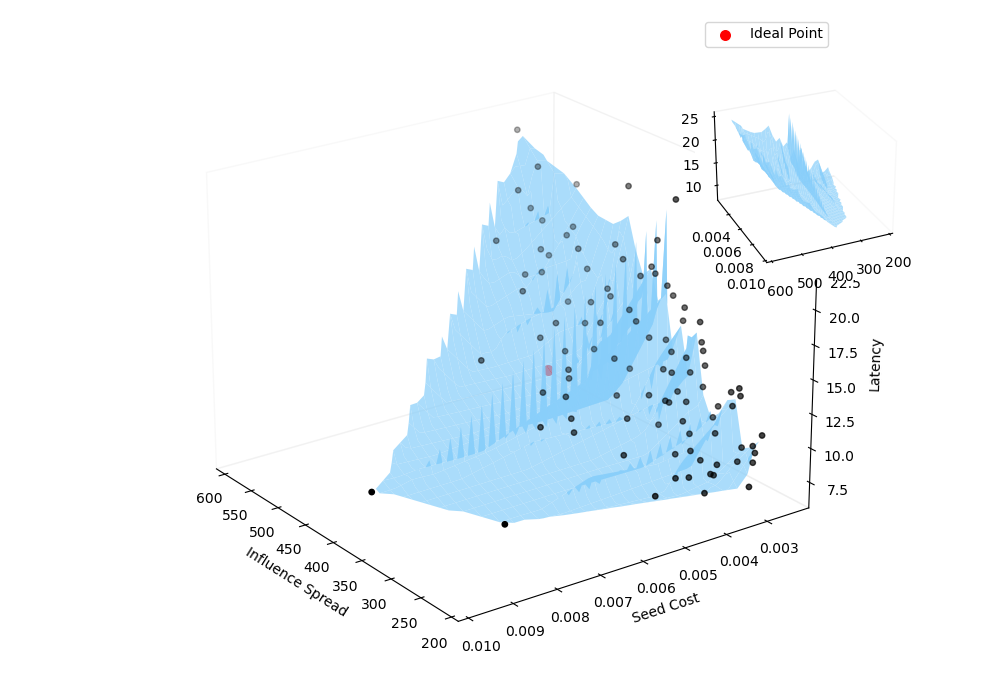}
         %\caption{6k}
         \label{paretofig:6000}}
     ~
     \subfigure[Wiki]{
         \centering
         \includegraphics[width=0.45\textwidth]{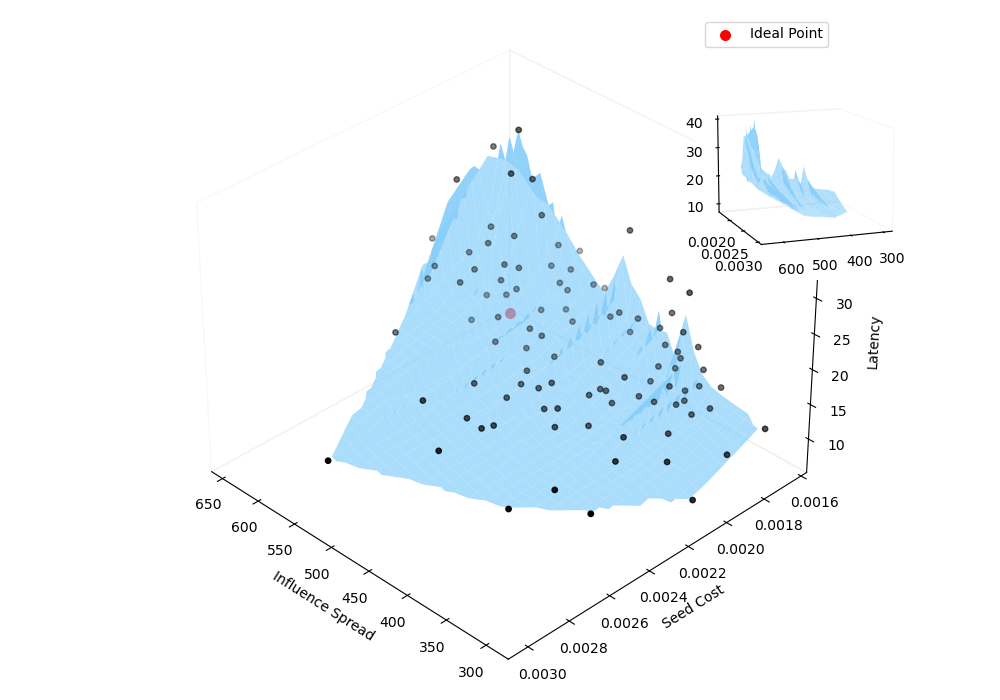}
         %\caption{7k}
         \label{paretofig:7000}}
     \caption{The Pareto-front of EVEA}
     \label{pareto}

\end{figure*}
\section{Experimental Results}

\begin{table*}[t]
\caption{Network Description}
\centering
\label{net}
\small
\begin{tabular}{llll}
\toprule
Network Name & Nodes$\#$ & Edges$\#$  & Description                               \\ 
\midrule
Facebook     & 4039  & 88234  & Social circles from Facebook (anonymized)         \\
GRQC         & 5242  & 14496  & Collaboration network of Arxiv General Relativity \\ 
Gnutella     & 6301 & 20777  & Gnutella peer-to-peer network from August 8 2002  \\ 
Wiki         & 7115 & 103689 & Wikipedia who-votes-on-whom network      \\ 
\bottomrule       
\end{tabular}
\end{table*}

This section presents the experimental results designed to evaluate the effectiveness and robustness of the proposed algorithm. We conduct experiments on four real-world network datasets and compare EVEA with several representative baselines. The results are analyzed from four perspectives: objective conflict demonstration, quantitative performance comparison, convergence behavior, and distributional analysis.

\subsection{Experimental Settings}
\begin{figure*}[t]
     \centering  
     \subfigure[Facebook]{
         \centering
         \includegraphics[width=0.45\textwidth]{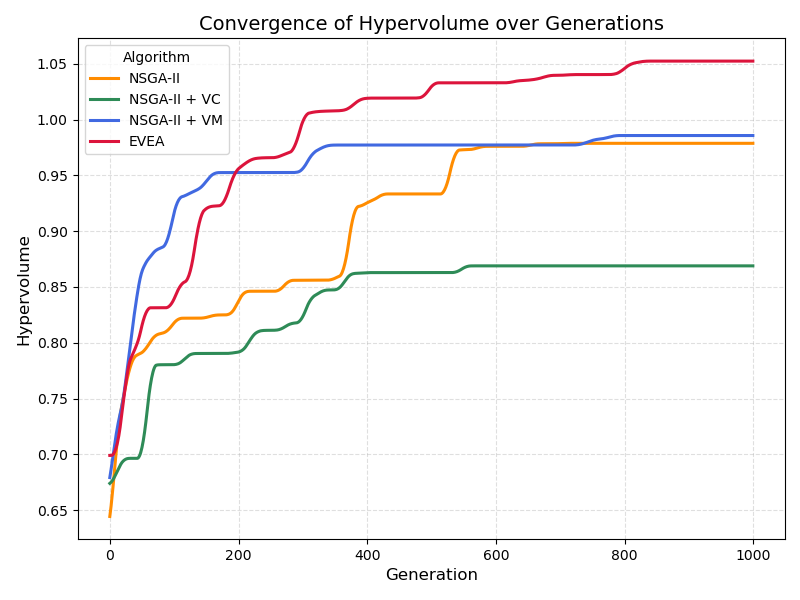}
         \label{fig:4000}}
     ~
     \subfigure[Grqc]{
         \centering
         \includegraphics[width=0.45\textwidth]{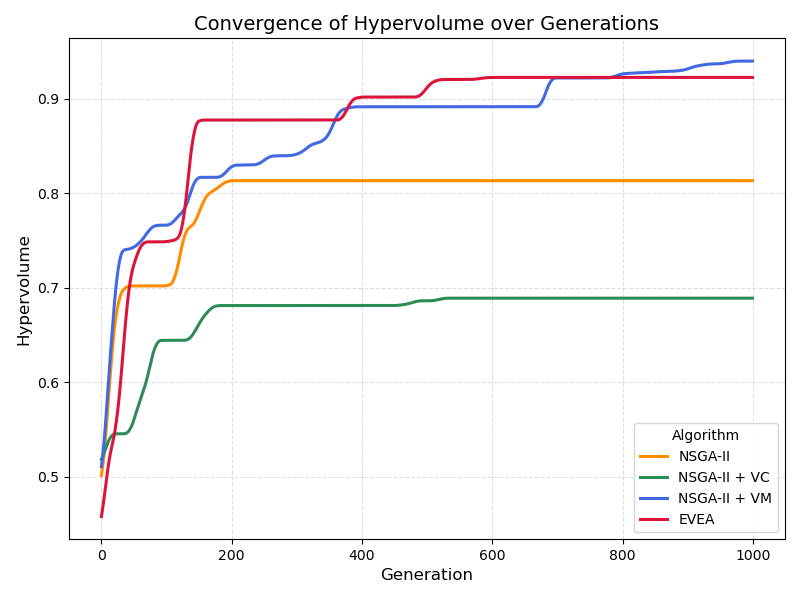}
         %\caption{5k}
         \label{fig:5000}}
     ~
     
     \subfigure[Gnutella]{
         \centering
         \includegraphics[width=0.45\textwidth]{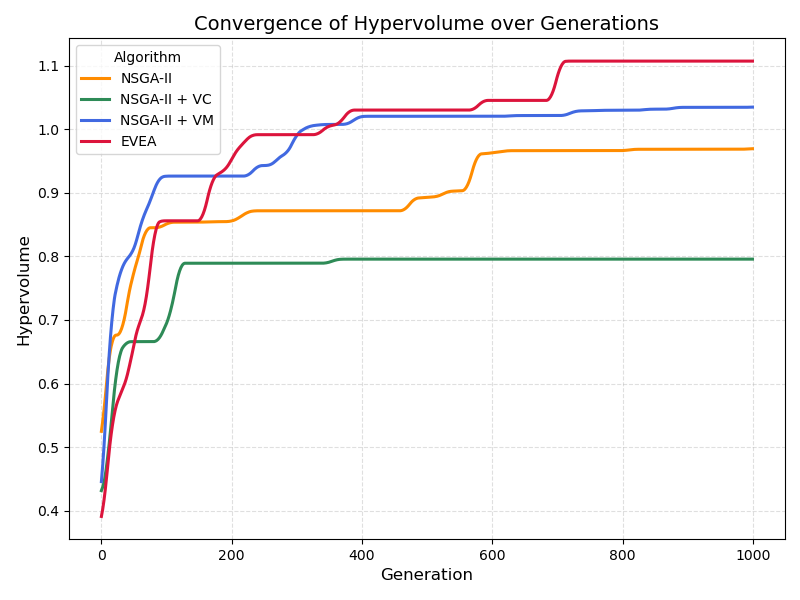}
         %\caption{6k}
         \label{fig:6000}}
     ~
     \subfigure[Wiki]{
         \centering
         \includegraphics[width=0.45\textwidth]{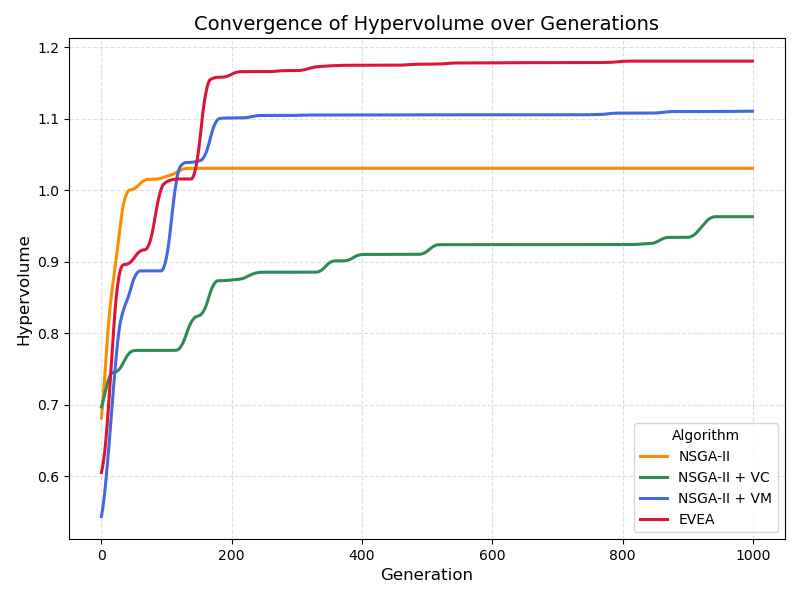}
         %\caption{7k}
         \label{fig:7000}}
     \caption{Convergence curves (HV vs. generations) of four algorithms across four networks.}
     \label{fig:convergence}

\end{figure*}

The experiments are conducted on four benchmark datasets - Facebook, GRQC, and Gnutella, all of which are obtained from the Stanford Network Analysis Project (SNAP) \cite{leskovec2016snap}. Each network exhibits different topological characteristics in terms of size, connectivity, and clustering. Table~\ref{net} summarizes the key statistics of these datasets.

We compare EVEA against three variants of NSGA-II: (1) standard NSGA-II \cite{deb2002fast}, (2) NSGA-II with variable-length crossover \cite{biswas2022improved}, and (3) NSGA-II with variable-length mutation \cite{genetti2024influence}. All algorithms are run for 1000 generations, and the population size is fixed across methods for fairness. The hypervolume (HV) indicator is used to assess the quality of the obtained Pareto fronts. We also report the average HV and standard deviation over multiple runs.

\subsection{Demonstrating Objective Conflicts}

To clearly illustrate the conflicting nature of the three objectives—maximizing influence spread, minimizing seed cost, and minimizing time—we present 3D Pareto front visualizations for four representative networks: Facebook, GRQC, Gnutella, and Wiki (Figure\ref{pareto}). In each case, the surface formed by the non-dominated solutions reveals a continuous and well-structured Pareto front in three-dimensional space. These surfaces are not trivial point clouds, but rather demonstrate smooth trade-off regions, indicating that the solutions are not concentrated at extreme points but instead spread across a spectrum of compromises among the objectives. The geometry of the Pareto surfaces, particularly the curvature and spread in each plot, further confirms that no single solution simultaneously optimizes all three objectives. Enhancing one objective, such as influence spread, inevitably results in a higher seed cost or longer time. Therefore, the presence of stable and interpretable Pareto surfaces across all four datasets validates the multi-objective nature of the problem and highlights the necessity of using a Pareto-based approach such as EVEA.

\subsection{Performance Comparison on Benchmark Networks}

We report the quantitative performance of EVEA and the baseline methods on four representative networks using the hypervolume (HV) indicator. The results, shown in Table~\ref{tab:hvresults}, reflect the average HV over multiple independent runs, with standard deviations provided in parentheses. The highest mean HV in each network is boldfaced, and statistically significant improvements by EVEA (based on the Wilcoxon signed-rank test at $p < 0.05$) are marked with an asterisk (*).

Overall, EVEA consistently outperforms all baseline methods across all four networks. On the \textbf{Facebook} network, EVEA achieves an HV of 1.0920, which is significantly higher than NSGA-II (0.9528), its crossover variant (0.8873), and the mutation variant (1.0137). This corresponds to relative improvements of 14.6\%, 23.1\%, and 7.7\%, respectively. A similar trend is observed on the \textbf{GRQC} network, where EVEA obtains the highest HV of 1.1023, compared to 0.9237 for NSGA-II, 0.7687 for the crossover variant, and 1.0090 for the mutation variant, reflecting margins of 19.3\%, 43.3\%, and 9.2\%.

On the \textbf{Gnutella} network, EVEA again yields the highest HV (1.1100), outperforming the second-best mutation variant (1.0628) by 4.4\% and clearly surpassing NSGA-II (0.9705) and its crossover-enhanced version (0.7969) by 14.3\% and 39.3\%, respectively. The advantage of EVEA is even more pronounced on the \textbf{Wiki} network, where it reaches an HV of 1.1659 with minimal variance, while the other methods remain notably behind—NSGA-II (1.0536), VM (1.1096), and VC (0.8987)—resulting in improvement margins of 10.6\%, 5.1\%, and 29.7\%.

These consistent improvements confirm the superiority of EVEA in discovering well-distributed and high-quality Pareto-optimal solutions across varying network topologies. The integration of embedding-aligned variable-length mutation and crossover operators enhances the flexibility of the search process, allowing the algorithm to adaptively explore diverse solution lengths and configurations with high robustness and effectiveness.

\begin{table*}[h]
\centering
\caption{Hypervolume (HV) Results with Standard Deviation on Four Networks. Values marked with $^*$ indicate statistical significance ($p < 0.05$) based on the Wilcoxon signed-rank test compared to all baselines.}
\label{tab:hvresults}
\begin{tabular}{lcccc}
\toprule
\textbf{Algorithm} & \textbf{Facebook} & \textbf{GRQC} & \textbf{Gnutella} & \textbf{Wiki} \\
\midrule
EVEA &  $\textbf{1.0920 $\pm$  0.0020}^*$ & $\textbf{1.1023 $\pm$ 0.0299}^*$ & $\textbf{1.1100 $\pm$ 0.0136}^*$ & $\textbf{1.1659 $\pm$ 0.0004}^*$ \\
NSGA-II &  $0.9528 \pm 0.0167$ & $0.9237 \pm 0.0299$ & $0.9705 \pm 0.0472 $ & $1.0536 \pm 0.0183 $ \\
NSGA-II + Crossover &  $0.8873 \pm 0.0269$ & $0.7687 \pm 0.0343$ & $0.7969 \pm 0.0304$ & $ 0.8987 \pm 0.0436 $ \\
NSGA-II + Mutation &  $1.0137 \pm 0.0244$ & $1.0090 \pm 0.0330$ & $ 1.0628 \pm 0.0196$ & $1.1096 \pm 0.0231 $ \\
\bottomrule
\end{tabular}
\end{table*}

\subsection{Convergence Analysis}

To assess the convergence behavior of the algorithms, we plot the evolution of hypervolume (HV) values over generations. The convergence curves, shown in Figure~\ref{fig:convergence}, illustrate the progression of Pareto front quality over time for each method across four benchmark networks.

EVEA consistently demonstrates both faster convergence and higher final HV values than all baseline algorithms. On the \textbf{Facebook} network, EVEA achieves over 90\% of its final HV within the first 50 generations, while NSGA-II and its variants require more than 100 generations to reach comparable levels. For example, at generation 50, EVEA reaches an HV of approximately 1.05, which is 11\% higher than that of NSGA-II (0.947). This trend continues, and by generation 100, EVEA has reached a stable HV near its maximum of 1.0920, clearly surpassing all baselines.

A similar pattern appears on the \textbf{GRQC} network, where EVEA steadily ascends to an HV of 1.1023, while NSGA-II converges more slowly and plateaus earlier. Notably, the HV of EVEA at generation 100 is 12.4\% higher than that of NSGA-II and nearly 44\% higher than that of NSGA-II + crossover. EVEA also maintains low variance throughout the run, highlighting its stability and robustness in complex search landscapes.

On the \textbf{Gnutella} network, EVEA reaches an HV of 1.09 by generation 100 and ultimately stabilizes at 1.1100. Compared to NSGA-II + crossover (0.7969), this represents a 39.3\% relative improvement. Meanwhile, NSGA-II + mutation performs better than the standard NSGA-II, but still trails EVEA by a significant margin.

Lastly, for the \textbf{Wiki} network, EVEA shows the most stable and rapid convergence among all methods, attaining an HV of 1.1659 with negligible variance ($\pm$ 0.0004). Other methods lag noticeably, with NSGA-II and its crossover variant showing both slower ascent and lower final values.

In summary, EVEA achieves superior convergence characteristics across all datasets—both in speed and solution quality—demonstrating its effectiveness in navigating the complex multi-objective search space and delivering high-quality, stable Pareto fronts in fewer generations.

\section{Conclusion}

In this paper, we proposed a novel formulation of the Influence Maximization problem that simultaneously optimizes three conflicting objectives: influence spread, seed cost, and time. This tri-objective formulation, referred to as IM-ICT, captures the practical trade-offs faced in real-world scenarios such as viral marketing and time-sensitive information campaigns. We demonstrated that these objectives are inherently in conflict through comprehensive empirical analysis on multiple real-world networks.

To solve the IM-ICT problem effectively, we developed an Embedding-aligned Variable-length Evolutionary Algorithm (EVEA). EVEA introduces two key innovations: a variable-length representation that enables flexible seed set sizes, and an embedding-guided crossover operator that preserves topological semantics while enhancing search diversity. These design choices allow EVEA to efficiently explore the solution space and maintain a well-distributed Pareto front.

Experimental results on four benchmark networks validated the superiority of EVEA over several baselines in terms of convergence speed, solution quality, and hypervolume performance. The 3D Pareto surfaces further illustrated that EVEA can uncover a wide spectrum of trade-off solutions, confirming the multi-objective nature of the problem and the importance of Pareto-based optimization.

For future work, we plan to extend our model to incorporate spatial-temporal constraints. In the current framework, time is modeled independently of node location. However, in many real-world applications, physical distance and network delay are closely coupled. Therefore, incorporating geospatial information and location-aware diffusion dynamics into influence maximization remains a promising and challenging research direction.

\bibliographystyle{unsrt}
\bibliography{article.bib} 
\end{document}